\documentstyle[preprint,prb,aps]{revtex}
\begin{document}
\draft
\preprint{}
%\twocolumn[\hsize\textwidth\columnwidth\hsize\csname @twocolumnfalse\
%endcsname

\title{Can quantum regression theorem be reconciled with
quantum fluctuation dissipation theorem ?}

\author{P. Shiktorov, E. Starikov, V. Gru\v zinskis}
\address{
Semiconductor Physics Institute,
A. Go\v stauto 11, 2600 Vilnius, Lithuania\\
e-mail: pavel@pav.pfi.lt}
 \author{L. Reggiani}
 \address{Dipartimento di Ingegneria dell' Innovazione,
 Istituto Nazionale di Fisica della Materia,
 Universit\`a di Lecce, Via Arnesano s/n, 73100 Lecce,Italy\\
e-mail: lino.reggiani@unile.it}
\date{\today}
\maketitle
\begin{abstract}
{
In the attempt to derive the regression theorem from the
fluctuation dissipation theorem several authors claim the 
violation of the former theorem in the quantum case. 
Here we pose the question:
does it exists a quantum fluctuation dissipation theorem (QFDT)
in its conventional interpretation ?
It is shown that the relation usually called as the QFDT
is the condition of detailed macroscopic energetic balance.
Following this interpretation the existing conflict between the
two theorems in the quantum case is removed.
}
\end{abstract}
\pacs{PACS numbers: 03.65.Ca, 05.30.-d, 05.40.+j}
\vskip2pc
\narrowtext

\section{Introduction}

Under thermal equilibrium conditions 
the  behavior of fluctuations of macroscopic 
observables of a physical system is governed by relationships which  
are formulated usually in terms 
of the regression theorem (the so called Onsager hypothesis [1]) 
and the  fluctuation-dissipation theorem [2-6]
(also known as the Nyquist relation).
The former pertains to the time domain and
states that the relaxation of a correlation of fluctuations
is described by the same law governing the  irreversible processes of
the observable quantity itself.
The latter pertains to the frequency domain and
interrelates in some universal way
the spectral characteristics of fluctuations and linear response 
(i.e. dissipation) of an observable of the physical system.
Often, the fluctuation dissipation theorem is written as:
\begin{equation}
S_{xx}(\omega)=g^s(\omega) Im\{\alpha_x(\omega) \}
\end{equation}
where $S_{xx}(\omega)$ is the spectral density of fluctuations
of the observable $x$, $Im\{\alpha_x(\omega)\}$ 
the imaginary part of the generalized susceptibility
which is responsible for the relaxation, and  
\begin{equation}
g^s(\omega)=\hbar coth({\hbar\omega\over 2kT})=
2\hbar(\overline{n}+{1\over 2}) 
\end{equation}
is the Planck relating factor
describing the average energy of the thermal radiation
of the field mode with frequency $\omega$.

In classical case $\hbar\rightarrow 0$,
the two theorems give the same description of the spectrum
of thermal fluctuations.
By contrast, within the more general quantum case
there appears a conflict between 
these two theorems since they
predict different behavior of $S_{xx}(\omega)$.
The quantum regression theorem (QRT) 
claims that the spectrum of fluctuations is determined
only by the permitted transitions between energetic states
of the system.
In turn, the quantum fluctuation dissipation theorem (QFDT) states, 
that besides the eigenfrequencies of
the system (as it required by the QRT)
in the spectrum of the fluctuations there exist additionally
the so called Matsubara frequencies, 
$\Omega_n=i{2\pi kT\over \hbar}n$, where $n=\pm 1,\ \pm 2,\ ...$.
The origin of these frequencies is related with the poles
of the Planck factor $g(\omega)$.

\par
This conflict is usually interpreted as a violation
of the QRT (see, for example, Refs. [7-9]).
In its most evident form such a violation is demonstrated in Ref. [9],
where the conclusion statement announced that 
"there is no quantum regression theorem".
The proof of the general character of such a statement
is based on the fact that the violation of  QRT
follows from QFDT.
However, a proof that QRT is valid independently of  QFDT
was given by Lax [10] on the  basis of
the  general principles of quantum statistics 
(see also Refs. [11-13]).
Since in Refs. [8,9] it is claimed that QFDT
contradicts the validity of  QRT, 
we argue  that the origin of such a conflict
is related with QFDT and its interpretation
(see also Ref. [14]).
\par
The aim of this article is to address this issue by   
considering the origin of such a conflict 
from a formal mathematical point of view.
\section{Formal derivation of the QFDT}
In the operator representation, the symmetrized correlation function 
of the fluctuations of an observable $x$ is written as [4-6]:
%
% Eq. 3
\begin{equation}
 C_{xx}(\tau)={1\over 2}
Tr\{\hat \rho_s[\hat x(\tau)\hat x(0)+\hat x(0)\hat x(\tau)]\}
\end{equation}
while the corresponding linear response function 
is given by Kubo formula [4-6]:
%
% Eq. 4
\begin{equation}
\alpha_x(\tau)={i\over \hbar}\Theta(\tau)
Tr\{\hat \rho_s[\hat x(\tau)\hat x(0)-\hat x(0)\hat x(\tau)]\}
\end{equation}
where $\hat \rho_s$ 
is the density operator which describes some
stationary state of the physical system under test characterized by
Hamiltonian $\hat H_s$, and $\Theta(\tau)$ is the unit step function.
\par
The linear response described by Eq. (4)
implies that the interaction between the system and the radiation
is determined by the semi-classical Hamiltonian $\hat V=-\hat x f(t)$,
where $f(t)$ is a classical force.
By performing the Fourier transform of Eqs. (3) and (4),
in accordance with ref. [6] one obtains:
%
% Eq. 5
\begin{equation}
\biggl [{ {S_{xx}(\omega)} \atop { \hbar Im\{\alpha_{x}(\omega)\} } }
\biggr ]
=
{1\over 2}
\biggl [ { {J_{xx}(\omega)+J_{xx}(-\omega)} 
\atop {J_{xx}(\omega) - J_{xx}(-\omega) } } \biggr ]
\end{equation}
where
\begin{equation}
J_{xx}(\omega)=\int_{-\infty}^{\infty}
Tr\{\hat \rho_s \hat x(\tau)\hat x(0)\}
e^{i\omega\tau}=
2\pi
\sum_{m,n} \rho_n |x_{mn}|^2
\delta (\omega_{mn}-\omega)
\end{equation}
is the spectral density corresponding to 
the one-directional in time correlation function,
$Tr\{\hat \rho_s \hat x(\tau)\hat x(0)\}$
and $J_{xx}(-\omega)$ corresponds to 
$Tr\{\hat \rho_s \hat x(0)\hat x(\tau)\}$,
$\rho_n$ is the probability to find the system
in the eigenstate with energy $E_n$,
$x_{mn}$ is the matrix representation of the operator $\hat x$
and $\omega_{mn}=(E_m-E_n)/\hbar$ is the frequency associated with
the transition between the energetic  states $E_m$ and $E_n$.
\par
Under thermal equilibrium 
$\hat \rho_s=exp(-\beta \hat H_s)/Tr\{exp(-\beta \hat H_s)\}$ 
with $\beta =1/(kT)$.
For the derivation of the QFDT an explicit
expression for $J_{xx}(\omega)$ is not necessary,
it is sufficient the existence of the quantum
spectral relation [6]:
\begin{equation}
J_{xx}(-\omega)=e^{-\beta\hbar\omega}J_{xx}(\omega)
\end{equation}
By substituting Eq. (7) into Eq. (5) one obtains:
\begin{equation}
\biggl [{ {S_{xx}(\omega)} \atop { \hbar Im\{\alpha_{x}(\omega)\} } }
\biggr ]
=
{1\over 2}
\biggl [ { {1+p(\omega) } 
\atop {1 - p(\omega) } } \biggr ]
J_{xx}(\omega)
\end{equation}
where $p(\omega)=exp({-\beta\hbar\omega})$.
According to Ref. [3-6], the QFDT is then obtained by excluding the factor 
$J_{xx}(\omega)$ which is
common to both $S_{xx}(\omega)$ and
$Im\{\alpha_{x}(\omega)\}$.
As a consequence of such a derivation,
the zeros of $Im\{\alpha_x(\omega)\}$ determined by the factor
$1-p(\omega)=0$
(see Eq. (8)) become the poles of the Planck factor in Eq. (1),
i.e., they originate the Matsubara frequencies
and, in turn, the QRT-QFDT conflict.
Thus, the poles of the Planck factor at the Matsubara frequencies
can not be considered as independent of the frequency dependence
of $Im\{\alpha_x(\omega)\}$ which, in accordance with Eq. (8),
has zeros at the same frequencies:
\begin{equation}
Im\{\alpha_x(\omega)\}\biggl |_{\omega=\Omega_n} =0
\end{equation}
Therefore, from a mathematical point of view, in this case
the right-hand side of Eq. (1) contains an indefinite form
of ${0\over 0}$-type.
\par
Now we pose the following open question: 
does the constraint given by Eq. (9)
merely represent the formal requirement necessary for a rigorous derivation
of the QFDT, or bring a proper physical meaning ?
\par
An attempt to answer this question is detailed in the following section. 
\section{Interrelation between the QFDT and the principles
of energy balance}
As physical model we shall consider a sufficiently large
isolated system subdivided into two subsystems. 
The first corresponds to some physical system under test,
the second represents the surrounding world.
In this case the total Hamiltonian can be written as:
$\hat H=\hat H_S+\hat H_T+\hat V$,
where $\hat H_S$ and $\hat H_T$ are Hamiltonians of subsystems, 
and $\hat V$ describes the interaction between these subsystems.
By using the standard procedure [15] 
to construct  the master equations
for the statistical operators of each subsystem,
$\hat \rho_i$ ($i=S,T$),
and assuming that the interaction is weak
one obtains the following equation for time variations of
the average energy $<\hat H_i>$ in the $i$-th subsystem:
%
% Eq. 10
\begin{equation}
{d \over dt} <\hat H_i> = {1\over \hbar^2}
\int_0^{\infty} Tr_{S+T}\{\hat \rho_S  \hat \rho_T
[[\hat H_i, \hat V], \hat V(-\tau)]\} d\tau
\end{equation}
where $<\hat H_i> = Tr_i\{\hat \rho_i \hat H_i \}$,
$\hat V(-\tau)=exp[-{i\over \hbar}(\hat H_S+\hat H_T)\tau]
\hat V exp[{i\over \hbar}(\hat H_S+\hat H_T)\tau]$.
\par
The energy exchange  described by Eq. (10)
satisfies the conservation law for the total energy of an isolated system
in the form corresponding to the assumption of a weak interaction:
${d\over dt}<\hat H_S+\hat H_T>=0$.
By assuming that the interaction Hamiltonian $\hat V=-\hat x \hat f$
is linear and factorized
with respect to the variables of both subsystems 
the matrix representation in the right-hand side 
of Eq. (10) takes the form:
%
% Eq. 11
\begin{equation}
{d \over dt} 
\biggl [
{<\hat H_S> \atop <\hat H_T>}
\biggr ]
= {\pi \over \hbar}
\sum_{m,n}^S \
\sum_{M,N}^T \
|x_{mn}|^2 |f_{MN}|^2
\biggl [
{ \omega_{mn}^S \atop \omega_{MN}^T}
\biggr ]
( \rho_m^S\rho_M^T - \rho_n^S\rho_N^T)
\delta(\omega_{mn}^S + \omega_{MN}^T)
\end{equation}
From Eq. (11) one directly obtains the condition of
the microscopic detailed energetic balance (MiDEB): 
%
% Eq. 12
\begin{equation}
\rho_m^S \rho_M^T = \rho_n^S \rho_N^T   \ \ \ \ \ or \ \ \ \ \
{\rho_m^S \over \rho_n^S} = {\rho_N^T \over \rho_M^T}
\end{equation}
which must be satisfied with respect to only those
energy states of both subsystems that
are directly involved in the interaction,
i.e.,  when $\omega_{mn}^S=\omega_{MN}^T$, $|x_{mn}|^2\neq 0$, 
$|f_{MN}|^2\neq 0$.
\par
Now, let us formulate the conditions of the energy balance
at the macroscopic level of description.
For this sake in Eq. (11) we replace  
the term $\delta(\omega_{mn}^S + \omega_{MN}^T)$ by
$\int \delta(\omega_{mn}^S -\omega) 
\delta (\omega_{MN}^T + \omega) d\omega $ and 
rewrite Eq. (11) 
by using the matrix representation of the asymmetric spectral density 
$J(\omega)$ given by Eq. (6):
%
% Eq. 13
\begin{equation}
{d \over dt} <\hat H_S> =
- {d \over dt} <\hat H_T>
 = {1 \over 4\pi  \hbar}
\int\omega [J_{xx}(-\omega)J_{ff}(\omega) 
- J_{xx}(\omega) J_{ff}(-\omega)] d \omega
\end{equation}
From Eq. (13) we obtain 
the condition of macroscopic detailed energy balance
(MaDEB) as:
%
% Eq. 14
\begin{equation}
J_{xx}(-\omega)J_{ff}(\omega) = J_{xx}(\omega) J_{ff}(-\omega)
\end{equation}
which requires to be fulfilled for any value of the 
current frequency $\omega$.
We notice that the condition given by Eq. (14) is not the only form 
which can be used to  express such a detailed balance.
For example, by using the definitions of 
$Im\{\alpha_x(\omega)\}$ and $S_{xx}(\omega)$ given by Eq. (5)
it is easy to show that Eq. (14) can be rewritten in an
equivalent form as:
%
% Eq. 15
\begin{equation}
Y_{f}(\omega)Im\{\alpha_x(\omega)\} = 
Y_{x}(\omega)Im\{\alpha_f(\omega)\} 
\end{equation}
where for $Y_{i}(\omega)$ 
one can use any of the three spectral densities, namely:
 $J_{ii}(\omega)$, $J_{ii}(-\omega)$ and $S_{ii}(\omega)$
($i=x,f$).
The equivalence of Eqs. (14) and (15) from the point of
view of the formal determination of 
$J(\omega)$, $S(\omega)$ and $Im\{\alpha(\omega)\}$
given by Eqs. (5) and (6)
does not mean their entire macroscopic equivalence.
This is related to the fact,
that with respect to Eq. (14), Eq. (15) 
involves additionally the
notion of the macroscopic function of linear response,
$\alpha_i(\omega)$, all properties of which cannot be obtained
from Eqs. (5) and (6) only.

Let us consider the conditions when the MiDEB given by Eq. (12)
and the MaDEB given by Eqs. (14) and (15)
can be considered as equivalent descriptions of the energy balance under
steady state.
In the following we shall use the formulation of balance conditions
through some ratio of characteristics of a subsystem
[see the right-hand side version of Eq.   (12)]
since such ratios sometimes are  universal functions
which are independent from the internal properties of the interacting
subsystems.
Let us rewrite Eq. (14) in the form:
%
% Eq. 16
\begin{equation}
{J_{xx}(-\omega)\over J_{xx}(\omega) } = 
{J_{ff}(-\omega)\over J_{ff}(\omega) }
\equiv p(\omega)
\ \ \ \ or \ \ \ \
{J_{ii}(-\omega)= p(\omega)  J_{ii}(\omega) }  
\end{equation}
where $p(\omega)$ is common for both subsystems factor which is
some single-valued function of the current frequency and it
satisfies the condition $p(-\omega)=p^{-1}(\omega)$.
It is easy to show that the equivalence of the energy balance
description given by Eq. (12) and (16)
is satisfied if $p(\omega)$ can be defined from the microscopic level as:
%
% Eq. 17
\begin{equation}
{\rho_m^S \over \rho_n^S} = {\rho_N^T \over \rho_M^T} = 
p(\omega)\biggl |_{\omega=\omega_{mn}=\omega_{NM}}
\end{equation}
Let us rewrite the MaDEB condition given by Eq. (15)
in a form analogous to Eq. (16)
\begin{equation}
{Y_{x}(\omega) \over Im\{\alpha_x(\omega)\} }= 
{Y_{f}(\omega) \over Im\{\alpha_f(\omega)\} }
\equiv g^Y(\omega)
\ \ \ \ or \ \ \ \ 
{Y_{i}(\omega) = g^Y(\omega) Im\{\alpha_i(\omega)\} }
\end{equation}
Here the function of current frequency $g^Y(\omega)$
will depend on which spectral density [i.e., the
symmetric $S(\omega)$ or asymmetric $J(\pm \omega)$]
is used to formulate the balance conditions.
The fulfillment of Eq. (16) allows us 
to represent the frequency dependence of 
$S_{ii}(\omega)$ and $Im\{\alpha_i(\omega)\}$
of both subsystems in a form
entirely analogous to thermal equilibrium [see Eq. (8)]
with the only difference that now  $p(\omega)$
is not necessarily given by the thermal value.
From the above it is easy to see that all the functions $g^Y(\omega)$
in Eq. (18) are determined by by the frequency dependence of
$p(\omega)$ only:
\begin{equation}
g^Y(\omega)={\hbar\over 1-p(\omega)}
\cases{
1  &,  $ \ for \ \ \ \ J(\omega)$ \cr
p(\omega) &, \ $for \ \ \ \ J(-\omega)$ \cr
1+p(\omega) &, \ $for \ \ \ \ S(\omega)$ \cr
}
\end{equation}
When  $p(\omega)$ is a universal function of frequency 
[e.g., in thermal equilibrium when
$p(\omega)=exp(-\beta\hbar\omega)$]
there is the  possibility to formulate the MaDEB conditions
in terms of expressions which relate in some universal way
the macro-characteristics of only one of the interacting subsystem
[see the right-hand side expressions in Eqs. (16) and (18)].
However, in so doing, it is easy to loose the physical meaning
of these expressions.

In thermal equilibrium, Eq. (18) allows the QFDT to be
given an alternative physical interpretation with respect to the
conventional one.
Indeed, by replacing in Eq. (18) $Y_{i}(\omega)$ with the symmetrical 
spectral density $S_{ii}(\omega)$  one obtains the MaDEB condition
[right-hand side expression in Eq. (18)] in a form which is identical
for both the subsystems and which coincides with the conventional form of 
the QFDT given by Eq. (1).
This allows us to conclude that the QFDT describes
the detailed energetic balance between the interacting
physical systems under thermal equilibrium.
As a consequence, the usual interpretation that the frequency dependences of
the Planck factor $g^s(\omega)$ [see Eq. (2)] and of 
the imaginary part of the generalized susceptibility
$Im\{\alpha_i(\omega)\}$ are independent 
(the source of the QRT-QFDT conflict)
is in contradiction with both the MiDEB and MaDEB principles.
When these principles are fulfilled,
the poles of $g^s(\omega)$ and the zeros of 
$Im\{\alpha_i(\omega)\}$ are determined by the same factor
$1-p(\omega)$ [see Eqs. (19) and (8), respectively].
The neglect of this property  when treating fluctuation phenomena
corresponds in essence to the violation of the MiDEB and MaDEB principles
and, hence, it implies the violation of the condition of stationarity 
at least at the Matsubara frequencies $\omega=\Omega_n$
where $p(\Omega_n)=1$.
\par
The formulation of the balance conditions
represented in terms of a ratio of macro-characteristics of 
one or of both subsystems can serve as a source
of incorrect interpretation of the MaDEB principle
and, as a consequence, can lead to the violation of the energy
conservation law for subsystems
with partially overlapping energetic spectra.
The typical example is the interaction between the harmonic oscillator
with eigenfrequency $\omega_s$ and the thermal bath
characterized by a quasi-continuous spectrum.
By using  for the matrix elements of oscillator the relations [15]:
$|x_{mn}|^2=(1/\omega_s)\gamma_s^2 \ \hbar (n+1)\delta_{m,n+1}$
and $\omega_{mn}^S=\omega_s(m-n)$,
from Eq. (6) one obtains asymmetric spectral densities
$J(\pm \omega)$ represented as:
%
% Eq. 20
\begin{equation}
\biggl [{ {J(-\omega)} \atop { J(\omega) } }
\biggr ]
=
2\pi{\gamma^2(\omega) \ \hbar \over \omega }
\biggl [ { {\overline{N}(\omega) } 
\atop {\overline{N}(\omega)+1 } } \biggr ]
\delta(\omega_s-\omega)
\end{equation}
where $\overline{N}(\omega_s)=\sum_{n=0}^{\infty}n\rho_n=
[exp(\beta\hbar\omega_s)-1]^{-1}$,
$\gamma^2(\omega_s)$ is the electro-dipole matrix element square
which can depend upon the oscillator frequency.
In Eq.(20), due to the properties of the $\delta$-function
we have performed the replacement 
usually used in such situations when
dependence of $\gamma^2(\omega_s)$
and $\overline{N}(\omega_s)$ on $\omega_s$ is replaced by
their dependence on the current frequency $\omega$.
In this way, from Eq. (20) one obtains:
%
% Eq. 21
\begin{equation}
{ {J(-\omega)} \over { J(\omega) } }=p(\omega)
={ {\overline{N}(\omega)} \over { \overline{N}(\omega) +1 } }
\end{equation}
which defines  in explicit form $p(\omega)$ in the whole frequency range,
while the spectrum of a single oscillator
is defined  at the frequency $\omega=\omega_s$ only,
i.e. it is described by a $\delta$-function which in Eq. (21) disappeared
from explicit consideration.
\par
By considering the thermal bath, also in this case it is used
the harmonic oscillator model which describes one mode
of the radiation field.
To obtain the spectral densities
$J(\pm \omega)$
for the thermal bath it is sufficient to perform in the right-hand
side of Eq. (20)
a summation over all the possible modes,
that is over subindex $s$.
In so doing, the dependence of 
$\gamma^2(\omega_s)$
and $\overline{N}(\omega_s)$ on $\omega_s$ 
but not on $\omega$ can remain.
By using the standard procedure
for systems with a quasi-continuous spectrum,
i.e., by replacing $\sum_s$ by $\int G(\omega_s)d\omega_s$,
where $G(\omega_s)$ is the density of the radiation mode with
frequency $\omega_s$ in the interval $d\omega_s$,
one obtains:
\begin{equation}
\biggl [{ {J(-\omega)} \atop { J(\omega) } }
\biggr ]
=
2\pi{G(\omega)\gamma^2(\omega)\hbar \over \omega }
\biggl [ { {\overline{N}(\omega) } 
\atop {\overline{N}(\omega)+1 } } \biggr ]
\end{equation}
From Eq. (22) one can again obtain Eq. (21),
which is now valid in the whole frequency range.
When a single oscillator interacts with the thermal bath,
the interaction takes place not in the whole frequency range,
as it would follow from Eqs. (21) or (18), but only at the
oscillator eigen-frequency, where the energy exchange 
is only possible.
\section{Conclusions and open questions}
Here we have shown that the QFDT expressed in its usual form of Eq. (1)
in essence represents the macroscopic principle of detailed energy
balance between a physical system and the surrounding world interacting
with it.
Such an interpretation, and the related restrictions
concerning the frequency dependence of the imaginary part of  
the generalized susceptibility of the system [see Eq. (9)], 
leads to two main consequences.
First, the announced conflict between QRT and QFDT related with
the Matsubara frequencies is reconciled.
Second, some conventional notions as macroscopic dissipation and relaxation
closely related to the QFDT should be revisited.
Indeed, these notions cannot be treated as internal characteristics
of one of the two interacting subsystems
which compose the total isolated system.

In the framework of the macroscopic approach where the energy dissipation
is described by $Im\{\alpha(\omega)\}$,
the energy conservation law under the energy balance
implies that the power dissipated by one of subsystems from another
must be returned back, i.e. it is equal to the power
dissipated by another subsystem from the first.
The form of the MaDEB given by Eq. (15) just express this
mathematically, by claiming that the spectra of the power 
dissipated by each subsystem from another are equal.
Therefore, with respect to one subsystem,
one side of Eq. (15) can be treated as the dissipated power
and another side as the power returned back, i.e.
as emitted power.
This is illustrated by the following example, which considers
the system interaction with thermal bath.
By using Eq. (22), the MaDEB conditions given by Eq. (15)
take the form:
%
% Eq. 23
\begin{equation}
2 \ \hbar\omega
\biggl [
{ { \overline{N}(\omega) +{1\over 2} \pm {1\over 2}} 
% \atop {\overline{N}(\omega) +1  } 
\atop {\overline{N}(\omega) +{1\over 2} } }
\biggr ]
Im\{\alpha_{x}(\omega)\}
=
\omega
\biggl [{  {J_{xx}(\pm\omega) } 
% \atop {J_{xx}(\omega) } 
\atop {S_{xx}(\omega)  } } 
\biggr ]
\end{equation}
The left-hand side of Eq. (23)
describes the power dissipated by the system 
when this power is provided by the thermal bath.
The right-hand side of Eq. (23) describes the power which the system 
returns back to the thermal bath.
If the only photon part 
$\hbar\omega\overline{N}(\omega)$, 
of the full energy of the field
is involved into the absorption process, 
then the returning  spectrum,
$J_{xx}(-\omega)$, corresponds to the spectrum of
the spontaneous emission processes occurring in the system.
This is easy shown by using the explicit form of $J(-\omega)$
given by  Eq. (6).
If now one reverses time, i.e. replace $\omega$ by $-\omega$,
the energy of the thermal bath
which is involved into dissipation process will be  
$\hbar\omega[\overline{N}(\omega)+1]$.
This means that  with respect to one of the interacting subsystems
the processes of energy exchange in equilibrium conditions
shows an asymmetry under time reversal.
If the system has absorbed the full energy of the field
together with the zero-field value,
$\hbar\omega[\overline{N}(\omega)+{1\over 2}]$
(the case of the QFDT), to fulfill MaDEB
it would be necessary that the system returns back the same full energy.
In this case, the emitted spectrum which returns back
would be described by the symmetric spectral density $S_{xx}(\omega)$.
Here the following question arises,
which spectrum of fluctuations and correspondingly which  
correlation function are experimentally measured: 
the asymmetric or symmetric ones ?
\par
In applications, the QFDT is often used to
describe the relaxation phenomena in systems interacting 
with a thermal bath.
In this interpretation as one of the formulation
of the MaDEB conditions there appears a question
about the correctness of such a use.
Indeed, if the relaxation process implies that the system is approaching
the stationary state, it means that the system is not in 
a stationary state.
It is evident that under nonstationary conditions
all the relations considered above loose their meaning,
and any attempt to use them will imply a violation
of the energy conservation law.
Thus, the QFDT can be applied for the description
of only those relaxation phenomena
which occur under  energy equilibrium conditions, as
described by the QFDT itself.
\par
The MaDEB conditions lead to another important property of
the relaxation processes:
the description of the relaxation in the interacting subsystems
which compose an isolated system cannot be considered
in independent way.
From the definition of the linear response function
$\alpha_i(\omega)$ it follows that
the spectral densities of the fluctuations in each of the subsystems
must be interrelated as
$S_{ii}(\omega)=|\alpha_i(\omega)|^2S_{jj}(\omega)$
where $i\neq j$.
Together with the MaDEB in the form given by Eq. (18) this leads
to additional restrictions put on the frequency dependence 
of $Im\{\alpha_i(\omega)\}$ pertaining to each of  the two subsystems:
%
% Eq. 24
\begin{equation}
Im\{\alpha_i^{-1}(\omega)\}= Im\{\alpha_j(\omega)\}
\end{equation}
where $Im\{\alpha_i^{-1}(\omega)\}=Im\{\alpha_i(\omega)\}
/|Im\{\alpha_i(\omega)\}|^2$
is the relaxation law in one of the subsystems,
which in accordance with Eq. (24)
is determined by the imaginary part of the generalized susceptibility
of the other subsystem.
For example, in the presence of an interaction of 
some systems with a thermal bath
characterized by the radiation mode density
$G(\omega)=\omega^2\pi^{-2}c^{-3}$ [15]
the relaxation in the system must be described by the law:
\begin{equation}
Im\{\alpha_i^{-1}(\omega)\}=
{1\over 2\hbar}
[J_{ff}(\omega)-J_{ff}(-\omega)]=
{\gamma^2(\omega) \over \pi c^3}\omega
\end{equation}
If the electrodipole matrix element square, $\gamma^2$,
is independent of $\omega$, one obtains the usual law
of viscous friction.

%
%\begin{equation}
%\end{equation}
%

%
\section*{Acknowledgments}
The financial support of the NATO
collaborative-linkage grant PST.CLG.977520, 
and the french-lithuanian bilateral
cooperation n. 12864 of french CNRS is acknowledged.


\begin{thebibliography}{99}

\bibitem{1}
L. Onsager, {\it Phys. Rev.} {\bf 38}, 2265 (1931).
%
\bibitem{2}
H. Nyquist, {\it Phys. Rev.} {\bf 32}, 110 (1928).
%
\bibitem{3}
H.B. Callen and T.A. Welton, {\it Phys. Rev.} {\bf 83}, 34 (1951).
%
\bibitem{4}
L. Landau and E. Lifshitz, {\it Statistical Physics}
(Addison Wesley Reading, Mass., 1974)
%
\bibitem{5}
R. Kubo, M. Toda, N. Hashitsume,
{\it Statistical Physics II} (Spriger-Verlag, Berlin, 1985)
%
\bibitem{6}
D.N. Zubarev, {\it Nonequilibium Statistical Thermodinamics},
(Nauka, Moscow, 1971)
%
\bibitem{7}
H. Grabert, {\it Z. Phys.} {\bf B 49}, 161 (1982).
%
\bibitem{8}
P. Tolkner,  {\it Ann. Phys. (N.Y.)} {\bf 167}, 380 (1986).
%
\bibitem{9}
G.W. Ford and R.F. O'Connell, {\it Phys. Rev. Lett.} {\bf 77}, 
798 (1996).
%
\bibitem{10}
M. Lax,  {\it Phys. Rev.} {\bf 129}, 2342 (1963).
%
\bibitem{11}
V.H. Louisell, {\it Quantum Statistical Properties of Radiation}
(Wiley, New York, 1973), Sec. 6.6.
%
\bibitem{12}
C. Cohen-Tannoudji, J. Dupont-Roc and G. Grinberg,
{\it Atom-Photon Interactions} (Wiley, New York, 1992), 
Chap. IV, p. 350.
%
\bibitem{13}
A. Mandel and E. Wolf, 
{\it Optical Coherence and Quantum Optics}
(Cambridge University Press, Cambridge, 1995), Sec. 17.1.
%
\bibitem{14}
P. Shiktorov, E. Starikov, V. Gru\v zinskis, L. Reggiani
L. Varani and J.C. Vaissi\`ere,
{\it cond-matt/0011420}, Nov. 24, 2000.
%
\bibitem{15}
%
P. Meystre, M. Sargent, {\it Elements of Quantum Optics}
(Springer-Verlag, 1991).


\end{thebibliography}
\end{document}